\documentclass{raa}

\usepackage{graphicx,times,amssymb,hyperref}        
\input{epsf.sty}                        
\input{psfig.sty}                       

\begin{document}

\title{Confirming the 115.5-day periodicity in the X-ray light curve of ULX NGC~5408 X-1}

   \volnopage{Vol.0 (200x) No.0, 000--000}      
   \setcounter{page}{1}          
\author{Xu~Han          \inst{1}
   \and Tao~An*         \inst{2,4}
   \and Jun-Yi~Wang     \inst{3,1}
   \and Ji-Ming~Lin     \inst{1}
   \and Ming-Jie~Xie    \inst{1}
   \and Hai-Guang~Xu    \inst{5}
   \and Xiao-Yu~Hong    \inst{2,4}
   \and Sandor Frey     \inst{6}
   }

   \institute{Key Laboratory of Cognitive Radio $\&$ Information Processing, the Ministry of Education, Guilin University of Electronic Technology, Guilin 541004, China; {\it wangjy@guet.edu.cn}\\
        \and
Shanghai Astronomical Observatory, CAS, Shanghai 200030, China;  {\it antao@shao.ac.cn}
       \and
School of Mathematics and Computing Science, Xiangtan University, Hunan 411105, China;
   \and
Key Laboratory of Radio Astronomy, CAS, Nanjing, 210008, China;
   \and
Department of Physics, Shanghai Jiao Tong University, Shanghai 200240, China;
   \and
F\"{O}MI Satellite Geodetic Observatory, P.O. Box 585, H-1592 Budapest, Hungary.
   }

   \date{Received~~2009 month day; accepted~~2009~~month day}

\abstract{
The Swift/XRT light curve of the ultraluminous X-ray (ULX) source NGC 5408 X-1 was re-analyzed with two new numerical approaches, Weighted Wavelet $Z$-transform (WWZ) and CLEANest, that are different from previous studies.
Both techniques detected a prominent periodicity with a time scale of $115.5\pm1.5$ days, in excellent agreement with the detection of the same periodicity first reported by Strohmayer (2009).
Monte Carlo simulation was employed to test the statisiticak confidence of the 115.5-day periodicity, yielding a statistical significance of $> 99.98 \%$ (or $>3.8\sigma$).
The robust detection of the 115.5-day quasi-periodic oscillations (QPOs), if it is due to the orbital motion of the binary, would infer a mass of a few thousand $M_\odot$ for the central black hole, implying an intermediate-mass black hole in NGC 5408 X-1.
}
   \authorrunning{Han, An, Wang et al.}            
   \titlerunning{Confirming the Periodicity in NGC~5408 X-1}  

   \maketitle
%
\section{Introduction}           
\label{sect:intro}

The presence of supermassive black holes (SMBHs, $10^{6}-10^{9}\, M_{\odot}$) in the hearts of active galactic nuclei (AGNs), and the stellar-mass black holes (BH) ($\sim$10 M$_\odot$) as the evolutionary remnants of massive stars has been widely accepted and also confirmed by extensive observations with various techniques. However the existence of intermediate-mass black holes (IMBHs, $10^2 - 10^4 M_\odot$) remains a puzzle. IMBHs are thought as the seeds of the SMBHs, the masses of which were subsequently built up either through the accretion of gas, or through hierarchical black hole mergers (reviewed by  \cite{Vol10}). Therefore discovery of IMBHs is of vital importance for understanding the dynamics of stellar clusters, formation of SMBHs and production of gravitational waves (\cite{MC04}).

The detection of a population of ultraluminous X-ray sources (ULXs, X-ray luminosity exceeding $10^{39}$~erg~s$^{-1}$) invokes an interpretation that they might be accretion-powered objects which harbour IMBHs (e.g. \cite{CM99}). However, the high X-ray luminosity may also result from relativistically beamed emission. In this regard, the central object is not necessary an IMBH (\cite{King01})
,  or the high X-ray luminosity may represent a super-Eddington accretion of stellar-mass black holes (\cite{Beg02}).
Alternatively some regular ULXs might represent an extension of low-mass X-ray binaries (LMXBs; \cite{Liu06}).
Determination of the mass of ULXs is crucial to distinguish between these rival interpretations of the physical nature of ULXs. The canonical methods of measuring dynamical mass that rely on direct measurement of the binary motion of the companion are often frustrated because the companion is usually too dim to be detected. On the other hand, indirect methods that are based on the fit of X-ray spectral energy distribution (e.g., \cite{Mak00}) and X-ray timing observations (e.g., \cite{Kaaret06}) also place constraints on the black hole mass .

NGC 5408 X-1 is one of the few ULXs showing variable X-ray flux (\cite{Sor04}). The source shows QPOs around 10--40 mHz observed with the XMM-Newton (\cite{Str07,DS12}). Recently Strohmayer (2009) claimed to discover QPOs in NGC~5408 X-1 with $P=115.5\pm4$ days from the Swift/XRT monitoring data. However Kaaret \& Feng (2009) analyzed the same data set (the only difference is that Strohmayer did exposure map corrections in the light curve) but did not detect significant periodicity, although they indeed found a clue of a 115-day periodicity.
In contrast with Galactic X-ray binaries whose characteristic variability period is of the order of days, it requires a rather long time span to search for periodicity of $\sim$100 days in NGC 5408 X-1. Except for M82 X-1 (\cite{KF07}), there are few other ULXs that have been monitored for such a long period.
The QPO of NGC 5408 X-1 is crucial for better constraining the physical nature of this ULX: if the 115.5-day periodicity is due to the orbital motion of the binary, that would provide an evidence for an IMBH in NGC 5408 X-1 (\cite{Str09b}). Both groups mainly relied on the periodicity analysis on the Lomb--Scargle (LS) periodogram (\cite{Lomb76,Sca81}). However the statisitical significance of the LS-technique-detected periodicities is still a controversial issue. The LS periodogram introduces a traditional confidence limit based on an exponential distribution $e^{-z}$ ($z$ is the highest peak), that is actually not accurate enough for random signals. Therefore additional time series analysis techniques are necessary to verify the detected periodicity.

In this Letter, we shall re-analyze the light curve of NGC 5408 X-1 to check the 115.5-day periodicity by using two other numerical techniques, the Weighted Wavelet Z-transform (WWZ) and the CLEANest methods which are more robust than the LS periodogram. The statistical significance of detected periodicities will also be evaluated with the Monte Carlo simulations.

\section{The algorithm} \label{sect:data}

Traditional methods which are used for searching for and identifying periodic fluctuations in time series are based on Fourier transform.
A modified Fourier-based periodogram method is called Date-Compensated Discrete Fourier Transform (DCDFT: \cite{Ferr81}) which models the observed data as a linear combination of three trial functions of constant $ 1(t)$, $cos(\omega t)$, $sin(\omega t)$.
Astronomical data are usually not evenly sampled. However Fourier transform of an unevenly spaced time series may often introduce a myriad of complications, resulting in altering the peak frequency slightly and changing the amplitude greatly, and even generating strong fake peaks.
In order to identifying true periodic signals out of a time series containing multiple periodicities, an updated algorithm named CLEANest (\cite{Fost95}) was developed.
The CLEANest technique successively subtracts the strongest peaks in the original power spectrum detected by the DCDFT, until the remaining strongest peak in the residual spectrum is statistically not significant.
These processes stop when the residual spectrum is considerably low and flat.
The advantage of the CLEANest technique is that
the significant frequencies involved would be refined at each iteration of the CLEANest procedure, thus the fake peaks induced by the sidelobe effect of the primary peaks are greatly reduced.

Compared to Fourier analysis method which represents only frequency analysis, the wavelet transform is remarkable for its localization properperty in both time and freqeuncy domains. The wavelet transform decomposes the signals into a combination of wavelet functions which can be scaled and shifted in time and frequency. The (quasi-)periodic behaviors of the signals can be identified through visual observations of the wavelet power as a function of translation (time) and dilation (frequency), and the characteristic frequency (or time scale) is determined by looking for peaks in the time-averaged wavelet power
(\cite{Gro89}).
For unevenly-sampled data, the fluctuations of the local number density may induce spurious high-frequency spikes and also result in fake time evolution behavior of the characteristic periods.
In order to solve these problems, Foster (1996) proposed a rescaled wavelet method called Weighted Wavelet $Z$-transform (WWZ).
Same with the canonical wavelet transforms, the WWZ is also a kind of function projection which projects the signals onto three trial functions: a constant function $ \varphi_{1}(t)=1(t)$, $\varphi_{2}(t)=cos(\omega(t-\tau))$, $\varphi_{3}(t)=sin(\omega(t-\tau))$, where $\omega$ is the dilation parameter, also called scale factor, and $\tau$ is the time shift.. But the difference is that the WWZ adopts statistical weights $\omega_{a}=exp[-c\omega^2(t_\alpha-\tau)^2]$ ($\alpha$=1, 2, 3) to the projection on the trial functions, where the constant $c$ determines how rapidly the wavelet decays.
The WWZ power is defined as $Z$-statistics, $WWZ=(N_{eff}-3)/2(V_x-V_y)$  (Foster 1996),
where $N_{eff}$ is the effective number of the data which represents the statistical data density, and $V_{x}$ and $V_{y}$ are the weighted variation of the data and the weighted variation of the model function, respectively.
In doing so, the WWZ follows the $F$-distribution with $N_{eff}-3$ and 2 degrees of freedom.
The CLEANest and WWZ technique have been widely used in studies of variable stars (e.g., \cite{Kiss99,Tem04}). Recently they are also successfully applied in X-ray timing analysis of AGNs and X-ray binaries (e.g., \cite{Esp08,Lac06}).

\begin{figure}
  \centering
  \includegraphics[width=0.9\textwidth]{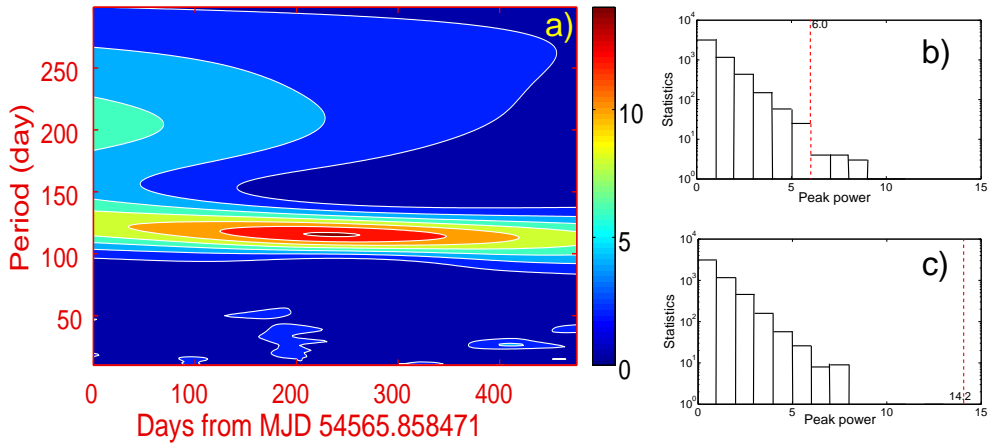}
\vspace{-5mm}
\caption{WWZ periodogram of NGC~5408 X-1 ({\it a}) and the statistical histograms of Monte-Carlo simulations of the 210-day periodicity ({\it b}) and of the 115.5-day periodicity ({\it c}).
The red dashed lines in (b) and (c) denote the peak powers of the periodic components in Fig. 1-a. To the right of the boundary line, the WWZ power of the simulated light curves is higher than the observed value; the ratio of the accumulated count of these light curves to the total test number (i.e., 5000) corresponds to a probability of the periodicity induced by the pure noise.
}
\label{fig:wwz}
\end{figure}

\section{Results}

The periodicity analysis results derived from the WWZ are presented in Figure \ref{fig:wwz}. The WWZ power is shown as a function of both the observing time ($x$-axis) and the testing period ($y$-axis) (Fig. \ref{fig:wwz}-a). The peaks in the WWZ power indicate the strength and duration of a characteristic periodicity. The majority of the WWZ power is dominated by the most prominent component whose peak marks a characteristic period of 115.5 days, and this periodic component spreads across the whole time span with a scattering of $\pm 1.5$ days. The amplitude of the 115.5-day component is 14.2, significantly higher than the secondary peak of 6.0, which corresponds to a charasteristic period of $\sim$210 days. The ridge line of the contours of the 210-day periodic component shows a large variation from 184 to 226 days.

Figure 2 shows the power spectra made from the Fourier transform analysis: the DCDFT (Fig. 2-a) and the CLEANest (Fig. 2-b).
In the DCDFT spectrum (Fig. 2-a), there are two peaks higher than the half of the maximal amplitude. The primary peak is located at a characteristic frequency of 0.00866 day$^{-1}$, corresponding to a period of 115.5 days.
The secondary peak falls in a relatively broad range from 194 to 236 days; the highest peak denotes a period of $\sim$210 days.

The periodograms derived from the WWZ and DCDFT show excellent consistency, and they are in agreement with the previous results based on the Lomb--Scargle periodogram analysis (\cite{Str09a}) as well: a distinct primary peak at a period of 115.5 days, and a scattered secondary at a period of $\sim$210 days. Note Strohmayer (2009) also detected the secondary peak at P2$\sim$210 days in his Figure 2, although the author did not remark on the secondary. The harmonic relation between two characteristic periods (frequency ratio $\sim$2:1) and the scattering appearance of the secondary are reminiscent that the 210-day periodicity might be an artifact due to the sidelobe of the primary.
Therefore we further run the CLEANest algorithm to check the reliability of the secondary periodic component. After subtracting the strongest 115.5-day peak, the residual CLEANest spectrum shown in Fig. 2-b becomes rather flat, and does not show any significant peak any more. That gives evidence that the 210-day peak detected in the WWZ and DCDFT spectrum is probably a fake signal which is contaminated by the nearby primary 115.5-day periodic component.

In order to further constrain the nature of the periodicities detected by the WWZ, we performed statisitical confidence tests with the Monte Carlo simulations (\cite{Lin85}).
First we created a series of 5000 light curves following Possion noise distribution to represent the random observational errors.
The simulated light curves had the same number of the sampling (113), temporal baseline (485 days), mean sampling rate ($\sim$14 days) and variance as the observed light curve.
The observed data were randomized in this way.
We then calculated the WWZ periodograms for each of the simulated light curves using the same parameters as we did for the observed light curve.
Next in the frequency channel correponding to the 115.5-day periodicity,  we searched for the periodograms in which the peak power exceeds 14.2, {\it i.e.}, the peak power of the 115.5-day periodicity, and counted the number of the light curves.
This accumulated number represents the occurrence that the 115.5-day periodicity can be generated by pure noise.
The ratio of this occurrence to the total testing number denotes a probability of fake detection of the periodicity due to the noise in the observed light curve.
The 210-day periodicity was tested following the same procedure.
The histogram of the Monte Carlo simulations are displayed  in Figure \ref{fig:wwz}-b (210-day) and \ref{fig:wwz}-c (115.5-day).
In Fig. 1-c, most of the power of the simulated light curves is lower than 8.0, there are only two exceptions higher than 10.0 but lower than 14.2 (the peak power derived from the observed data), placing a lower limit of the confidence level of 99.98\% (or $>3.8\sigma$).
Compared with the observed WWZ periodogram in Fig. 1-a, the simulations provide a unambiguous evidence for the detection of the 115.5-day periodicity across the whole observing period.
In Fig. 1-b, there are 13 simulated light curves that may generate peak powers of the 210-day periodicity higher than that from the observed data, setting a upper limit of the statisitical confidence of 99.74\% ($\lesssim3\sigma$).

\begin{minipage}{0.6\textwidth}
  \includegraphics[width=0.8\textwidth]{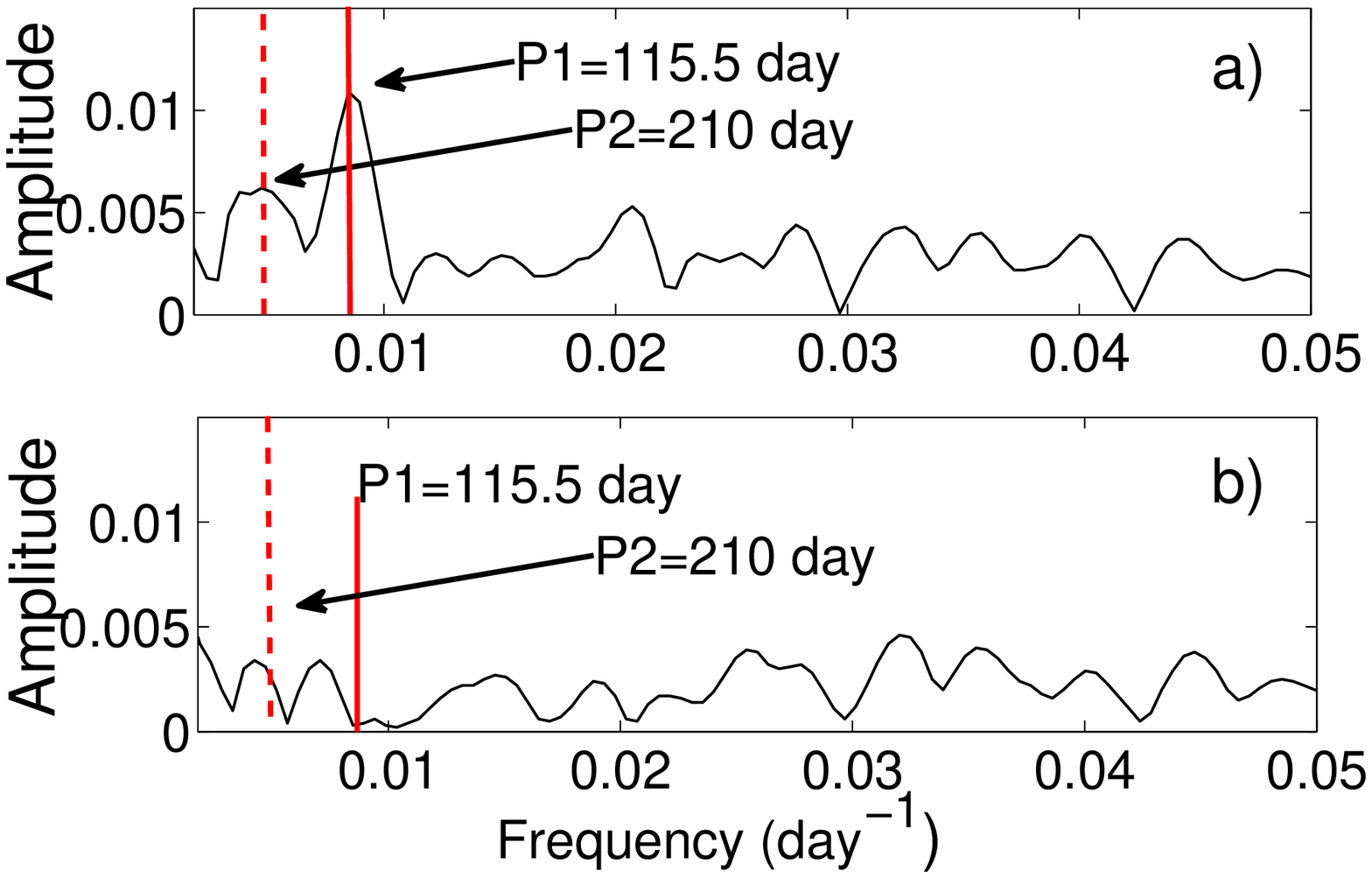}
\label{fig:cleanest}
\end{minipage}
\begin{minipage}{0.3\textwidth}
{\bf Fig. 2}  The DCDFT ({\it Top}) and CLEANest ({\it Bottom}) periodograms of NGC~5408~X-1. The red solid line indicates the 115.5-day periodic component.
The red dashed line indicates the fake 210-day period.
\end{minipage}

\section{DISCUSSION}
\label{sect:discussion}

In the present work, two numerical techniques, the WWZ and the CLEANest, were used to search for the periodicities in the Swift/XRT light curve of NGC~5408~X-1.
Both WWZ and DCDFT periodograms detected a distinct component across the whole time span correponding to a 115.5-day period, in good agreement with  previous works (\cite{Str09a}).
Monte Carlo simulation was employed to test the statisitical significance of the 115.5-day periodicity.
The simulations indicate a high confidence level of $>$99.98\%, excluding the possibility that the periodicity is generated by the random noise.
All the evidence points to a conclusion that the X-ray light curve of NGC~5248~X-1 intrinsically displays QPOs with a time scale of 115.5 days.

A straightforward interpretation of the 115.5-day periodicity is the orbital period of the binary system, resembling with the QPOs detected in Galactic BH binaries such as XTE J1550-564 and GRS 1915+105 (\cite{Str09a} and references therein).
The difference between NGC~5408~X-1 and the Galactic BH binaries is that the time scale in NGC 5408 X-1 is longer by a factor $\sim$100, inferring a larger dynamic mass of the central BH.
If the low-frequency QPO in NGC 5408 X-1 is analogous to Galactic black hole binaries in physical nature, the mass of the NGC 5408 X-1 black hole might be estimated from the Galactic BH mass and the scaling factor of the QPO frequencies. That gives an estimated mass of NGC 5408 X-1 BH ranging from 1000 M$_\odot$ to 9000 M$_\odot$ (\cite{Str09b}).

In addition to the orbital motion scenario, there are other arguments for this QPO in NGC 5408 X-1. Foster et al. (2010) claimed that the 115.5-day QPO is probably a manifestion of jet precession resembling SS~433, and is thus superorbital in nature. Therefore NGC 5408 X-1 might in fact be a large stellar-mass BH accreting with a super-Eddington rate to account for the high X-ray luminosity (\cite{Mid11}). Similarly, Soria (2007) proposed that the accretion properties of ULXs are different from stellar-mass BHs, and low-frequency milli-Hz QPOs cannot be used as an indicator of the high black hole mass. In their model, the ULXs can be 50--100 M$_\odot$-mass BHs with high accretion rate.

Continuing X-ray timing observations and energy spectra studies are necessary to discriminate between the scenarios for explaining the 115.5-day periodicity.
If further observations can prove that the 115.5-day periodicity is indeed related to the orbital motion of the binary system, NGC 5408 X-1 would be one of the few IMBH candidates in extragalaxies.

\begin{acknowledgements}
The authors thank the anonymous referee for quick response and helpful comments. This work is partly supported by the 973 Program of China (2009CB24900, 2012CB821800), the National Natural Science Foundation of China (61261017), the Strategic Priority Research Program of the CAS (XDA04060700), the China-Hungary Exchange Program of the CAS, and the Hungarian Scientific Research Fund (OTKA K104529).
\end{acknowledgements}

\end{document}